\documentclass[useAMS,usenatbib]{mnras}
\usepackage{amsmath}
\usepackage{amssymb}
\usepackage{hhline}
\usepackage{graphicx}
\usepackage{color}

\usepackage{newtxtext,newtxmath}        
\usepackage[T1]{fontenc}                
\usepackage{ae,aecompl}
\usepackage{natbib}
\newcommand{\be}{\begin{equation}}
\newcommand{\ee}{\end{equation}}

\newcommand{\SigSFRunits}{M_{\sun}\ \rm{kpc}^{-2}\ \rm{yr}^{-1}}
\newcommand{\e}{\times10^}

\begin{document}
\title[Feeding cosmic star formation]{Feeding cosmic star formation: Exploring high-redshift molecular gas with CO intensity mapping}

\author[Patrick C. Breysse and Mubdi Rahman]{Patrick C. Breysse$^{1}$\thanks{pbreyss3@jhu.edu} and Mubdi Rahman$^{1}$ \\
$^{1}$ Department of Physics and Astronomy, Johns Hopkins University, Baltimore, MD 21218 USA}

\maketitle
\label{firstpage}

\begin{abstract}
The study of molecular gas is crucial for understanding star formation, feedback, and the broader ecosystem of a galaxy as a whole. However, we have limited understanding of its physics and distribution in all but the nearest galaxies.  We present a new technique for studying the composition and distribution of molecular gas in high-redshift galaxies inaccessible to existing methods.  Our proposed approach is an extension of carbon monoxide intensity mapping methods, which have garnered significant experimental interest in recent years.  These intensity mapping surveys target the 115 GHz $^{12}$CO (1-0) line, but also contain emission from the substantially fainter 110 GHz $^{13}$CO (1-0) transition.  The method leverages the information contained in the $^{13}$CO line by cross-correlating pairs of frequency channels in an intensity mapping survey.  Since $^{13}$CO is emitted from the same medium as the $^{12}$CO, but saturates at a much higher column density, this cross-correlation provides valuable information about both the gas density distribution and isotopologue ratio, inaccessible from the $^{12}$CO alone.  Using a simple model of these molecular emission lines, we show that a future intensity mapping survey can constrain the abundance ratio of these two species and the fraction of emission from optically thick regions to order $\sim30\%$.  These measurements cannot be made by traditional CO observations, and consequently the proposed method will provide unique insight into the physics of star formation, feedback, and galactic ecology at high redshifts.
\end{abstract}


\section{Introduction}

Molecular gas provides the fuel and composes the environment from which stars form across cosmic time. Its abundance is directly related to the rate at which stars form \citep{2008AJ....136.2846B,2009AJ....137.4670L}, and its direct measurement provides insight into the efficiency through which this process takes place \citep{2011ApJ...729..133M}. 

Most available observations of molecular gas come from the Milky Way and other nearby galaxies.  However, recent years have seen an explosion of information about the high redshift universe, from precise measurements of the cosmic microwave background \citep{Planck2015} to large galaxy surveys such as the Sloan Digital Sky Survey \citep{Eisenstein2011}.  A consensus picture of the cosmic star formation history has emerged, with star formation rates gradually rising from the end of reionization to a peak around $2 \lesssim z \lesssim 3$ before beginning to decline \citep{Madau2014}.  Unfortunately, our understanding of molecular gas at these redshifts is still quite limited; direct molecular emission has only been measured for a small selection of sources at these redshifts, limited to the most luminous of far-infrared and submillimeter galaxies \citep{2005ARA&A..43..677S}. A full picture of high-redshift molecular gas and its relationship to star formation requires a more systematic approach.  

The rotational transitions of carbon monoxide molecules have long been some of the most potent tracers of the molecular gas phase of the interstellar medium.  H$_2$, though a much more common molecule, is not easily excited in typical molecular cloud conditions.  Reasons for this include the fact that H$_2$ lacks an electric dipole moment and has a low mass.  These properties raise the temperature required to excite the molecule out of the range of all but the most extreme molecular clouds.  CO, on the other hand, is easily excited.  Combined with millimeter emission wavelengths which fall into a relatively transparent atmospheric window, this has led to a great deal of study of CO and its dynamics (See reviews by \citet{Kennicutt2012}, \citet{Bolatto2013}, \citet{Carilli2013}, \citet{Heyer2015}, and others).   

However, there are drawbacks to using $^{12}$CO, the most common CO isotopologue, alone as a molecular gas tracer.  Since these rotational lines are strong, they saturate at relatively low column densities compared to other lines, becoming optically thick relatively soon after densities rise sufficiently to support CO formation in the first place \citep{Bolatto2013}.  Thus it is a common practice to combine observations of $^{12}$CO with those of other, fainter lines which can penetrate further into molecular clouds.  $^{13}$CO, a stable isotopologue of the usual $^{12}$CO molecule, is commonly used to gain information about denser material in environments ranging from protoplanetary disks \citep{Miotello2014} to giant molecular clouds (see, for example, \citet{Heyer1996}).  $^{13}$C has an abundance on the order of a few percent of that of $^{12}$C \citep[see, for example][]{Binney1998}, and the relative abundances of the two CO isotopologues is similar, up to corrections due to chemical fractionation and selective photodissociation \citep{Wilson1999}.  Combining measurements of $^{12}$CO and $^{13}$CO for a single system facilitates the study of molecular gas across a wider density range \citep{Pineda2010}.

At cosmological distances, the $^{12}$CO emission lines can only be observed in very bright sources, and $^{13}$CO observations are limited only to extreme systems such as the strongly lensed Cloverleaf quasar \citep{Henkel2010}.  The difficulty of detecting these lines severely limits what can be learned from traditional galaxy surveys at these redshifts.  Even in these very bright sources, most radio surveys lack the resolution needed to probe sub-galactic scales, preventing the study of gas distributions within any individual galaxy.

Intensity mapping is a method that probes the entire galaxy population, including the large numbers of faint sources inaccessible to traditional surveys.  These experiments seek to measure the intensity fluctuations of an emission line on scales large compared to galaxies.  While no source is detected individually, this process enables the statistical measurement of spectral lines using all photons emitted within the targeted volume.  In this paper, we will demonstrate how CO intensity maps can measure the detailed molecular gas composition and distribution at high redshifts.  These measurements will help unveil how star formation is fed in distant galaxies, systems that are all but invisible to current techniques.

For this paper, we will assume a simple $\Lambda$CDM cosmology with $(\Omega_m,\Omega_\Lambda,h,\sigma_8,n_s)=[0.27,0.73,0.7,0.8,0.96]$.  Section 2 provides a brief overview of intensity mapping and its application to high redshift astrophysics, along with a qualitative description of our proposed methods.  A more formal treatment follows in Section 3.  In Section 4, we forecast the detectability and constraining power of this method, with detailed discussion in Section 5.  We conclude in Section 6.

\section{Intensity Mapping}
First proposed by \citet{Suginohara1999}, intensity mapping experiments are now being planned for a wide variety of spectral lines.  The most popular line is the 21 cm spin-flip transition in neutral hydrogen (e.g. \citet{vanHaarlem2013,Bandura2014,Neben2016}).  Other lines being targeted include the 158 $\mu$m [CII] fine-structure line \citep{Gong2012,Crites2014,Yue2015}, and the Lyman-$\alpha$ line \citep{Silva2013,Pullen2014,Gong2014,Comaschi2016}.

Most interesting for the study of molecular gas though are planned intensity mapping experiments targeting the 115 GHz $^{12}$CO(1-0) line.  The CO intensity mapping signal was first modeled by \citet{Righi2008} as a possible CMB foreground, before being recognized as an interesting observable in its own right \citep{Lidz2011,Pullen2013,Mashian2015,Breysse2014,Breysse2016}.  The current experimental efforts in CO intensity mapping are primarily targeted at $2 \lesssim z  \lesssim 3$.  This is a particularly interesting redshift range as it is near the peak of the cosmic star formation rate.   The CO Power Spectrum Survey (COPSS), a high-resolution interferometric survey aiming to detect CO fluctuations on relatively small scales, recently published the first tentative detection of the CO signal at these redshifts \citep{Keating2015,Keating2016}.  A second experiment known as the CO Mapping Array Pathfinder (COMAP) is currently under construction \citep{Li2016}.  COMAP is a single-dish experiment which aims for larger scales than COPSS.

$^{12}$CO intensity maps produced by these experiments will reveal a large amount of information about high-redshift molecular gas.   however they will be limited by the same saturation effects seen in local-universe CO observations.  But as in the local universe, we can use multiple lines to aid our understanding.  Every frequency channel of an intensity mapping survey will contain emission from many lines besides the targeted line, coming from sources at a variety of different redshifts.  In some cases these interloper lines may be bright enough to dominate over the desired target line, which will necessitate some means of cleaning out the unwanted emission \citep[e.g.,][]{Breysse2015, Gong2014, Lidz2016}.  However, since one naturally seeks to measure the brightest, most accessible lines, the majority of interloper lines in any survey will be significantly fainter than the target line.  These faint lines may still contain useful information that can be extracted through the use of cross correlation.  Just as in local observations, $^{13}$CO provides a powerful complement to $^{12}$CO for this purpose.

\subsection{Leveraging Isotopologues}
Any $^{12}$CO survey will also contain $^{13}$CO in an overlapping cosmological volume.  Though $^{13}$CO is too faint and only marginally separated in frequency in comparison to $^{12}$CO to ever be detected by itself, by cross-correlating the proper frequency bands in a CO survey, we can measure the correlation between the intensities of these lines.  Since the lines come from the same population of galaxies, this cross-correlation depends on the relative abundances of the two carbon species and the optical depths of the CO lines.  Thus by performing this cross-correlation, we can make statements about the distribution of gas densities in our galaxy population, despite having no information about any individual galaxy or molecular cloud.  This measurement cannot easily be made using any other method, yet it is available without any extra observation from any CO intensity mapping survey.

\begin{figure}
\centering
\includegraphics[width=\columnwidth]{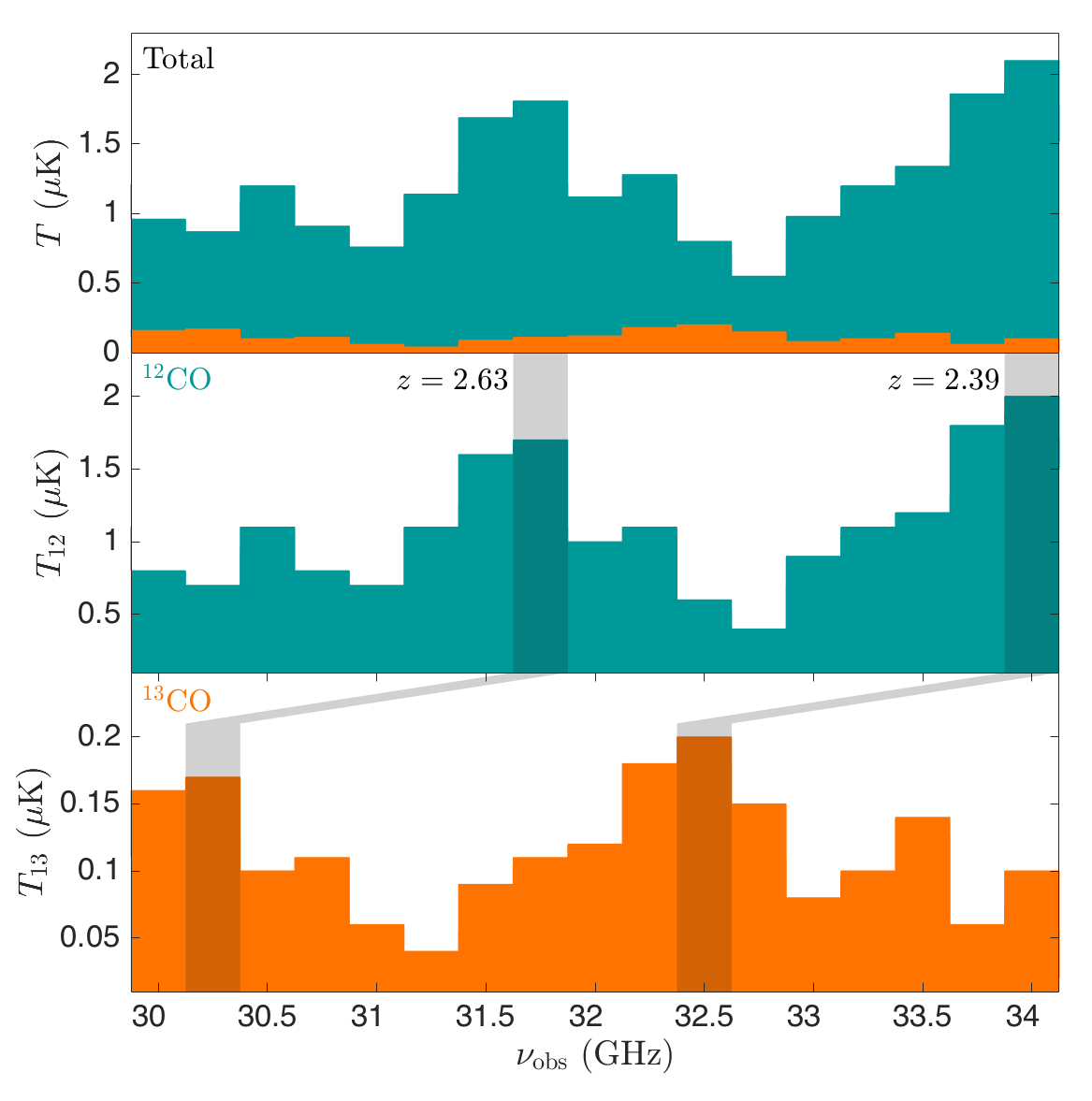}
\caption{A schematic view of the contributions from $^{12}$CO and $^{13}$CO to a hypothetical intensity mapping survey.  The top panel shows the total observed intensity in each frequency bin assuming that the observed $^{13}$CO intensity from all galaxies is 10\% of the $^{12}$CO intensity.  The middle and bottom panels show the contribution to the total signal from $^{12}$CO and $^{13}$CO emission respectively.  The shaded regions in these panels highlight emission that comes from the same slice of physical space.}
\label{Visual}
\end{figure}

Consider a CO intensity mapping survey similar to the planned COMAP experiment described in Table 2 of \citet{Li2016} which would cover a frequency range from 30 to 34 GHz.  This corresponds to $z=2.39$ to 2.84 in $^{12}$CO and $z=2.24$ to 2.67 in $^{13}$CO.  Figure \ref{Visual} shows a schematic view of what the contributions to a survey from these two lines might look like.  A given set of galaxies will emit both $^{12}$CO and $^{13}$CO lines at a given position in physical space.  The two lines are then redshifted to different bands in frequency space as shown in the bottom two panels, then added together to produce the observed signal in the top panel.  For illustration purposes, Figure \ref{Visual} assumes that the observed $^{13}$CO intensity from all galaxies is 10\% of the $^{12}$CO intensity.  

Even with a somewhat generous 10\% intensity ratio, the contribution from $^{13}$CO to the total intensity is small, and will therefore be virtually impossible to isolate in any individual band.  However, looking at the bottom two panels of Figure \ref{Visual} demonstrates that the two lines trace the same structure, just shifted in frequency.  At $z=2.6$, this shift in observed frequency is $\Delta\nu_{12/13}=1.4$ GHz.  The shaded bands show the two peaks which appear at the same location in physical space.  This means that we should expect significant correlation between bands separated by $\Delta\nu_{12/13}$.  By cross-correlating said bands we can compare the emission in both isotopologues coming from the same set of sources.

What we will observe in this cross-correlation depends on the details of the molecular gas properties in the surveyed volume.  Obviously we expect the signal to depend on the relative amounts of $^{12}$C and $^{13}$C, which depends on the nucleosynthesis history up to the observed redshift.  The observed signal will also depend on the density of the emitting molecular clouds.  Since $^{13}$C is significantly less abundant than $^{12}$C, we expect the $^{12}$CO line to become optically thick well before the the $^{13}$CO line does.    If much of the CO molecules in a galaxy are found in very dense regions, this effect can significantly alter the observed intensities.

It is relatively straightforward to qualitatively predict what the observed intensity ratio $I_{13}/I_{12}$ between the lines will be in two extreme cases.  We will ignore for now small differences between the emission and absorption properties of the two molecules.  In the limit of very diffuse gas, where all of the $^{12}$CO emission is optically thin, the only thing that matters is the relative abundance of the two species and we expect $I_{13}/I_{12}$ to approach the abundance ratio of the two isotopologues.  In the opposite limit, where the emission comes from a very dense region and both lines are saturated, we expect $I_{13}/I_{12}$ to approach unity, since each photon will undergo a large enough number of emission and absorption events to wash out any dependence on the relative abundances.  These results depend only on basic radiative transfer physics, and are independent of how the gas is modeled.

In real galaxies, we expect there to be contributions from both optically thin and optically thick regions.  By cross-correlating frequency bands separated by the frequency difference between the two isotopologues, we can constrain both the abundance ratio of the two species and the amount of emission that comes from dense, optically thick regions.  Both of these properties depend on both the instantaneous star formation rate and the cumulative star formation history at a given redshift.  Below, we will derive this intensity ratio as a function of these quantities in a more quantitative fashion.

\section{Formalism}
In this section we will present a general formalism for predicting the cross power spectrum of $^{12}$CO and $^{13}$CO emission in an intensity map.  This method should be applicable regardless of the exact model used to predict the molecular gas properties of a galaxy distribution.

\subsection{$^{12}$CO/$^{13}$CO Intensity Ratios}
We derive the intensity ratio of the two CO lines for a given molecular cloud and its dependence on gas properties informed by \citet{Pineda2010}. For the purpose of this work, we assume that the different isotopologues of CO are uniformly mixed throughout the medium, and consequently, both molecules will have identical excitation environments at any given line of sight, including densities and temperatures. Consider a generic emission line coming from a line of sight through molecular material with total optical depth $\tau_r$.  If there is no background intensity and the line profile is narrow compared to the frequency bandwidth of the observing instrument, then the observed intensity in a given band will be
\be
I = \int_0^{\tau_r} \frac{j(\tau)}{\kappa(\tau)}e^{-\tau}d\tau,
\label{Transfer}
\ee
where $j(\tau)$ is the emissivity and $\kappa(\tau)=n\sigma$ is the absorption coefficient for a molecule with number density $n$ and absorption cross section $\sigma$.  Because our observations come from only a fairly narrow redshift range, it is reasonable to assume that there is no significant effect from any background sources.  This also means that the probability of seeing emission from two galaxies in the same frequency band along the same line of sight is negligible. 

From \citet{Spitzer} the emissivity is
\be
j=\frac{h\nu}{4\pi}n_uA_{ul},
\ee
where $A_{ul}$ is the Einstein $A$ coefficient for the transition, $\nu$ is the rest frame emission frequency, and $n_u$ is the number density of molecules in the upper state.  We can rewrite this in terms of the total number density $n$ as
\be
j=\frac{3h^{2}\nu^2A_{ul}}{8\pi k_BT_{\rm{ex}}}ne^{-h\nu/k_BT_{\rm{ex}}},
\ee
where $T_{\rm{ex}}$ is the excitation temperature of the molecule and we have approximated the partition function as $Z\approx2kT/h\nu$.  This approximation is valid when $kT_{\rm{ex}}\gg2h\nu$ \citep{Pineda2010}, and thus should hold in typical molecular cloud conditions.  

If we change the integration variable in equation (\ref{Transfer}) to column density we have
\be
I=\frac{3h^{2}\nu^2A_{ul}}{8\pi kT_{\rm{ex}}\sigma}e^{-h\nu/k_BT_{\rm{ex}}}\left(1-e^{-N_r^{l}\sigma}\right),
\ee
where $N_r^{l}$ is the total column density of molecules in the lower state, and we have assumed that the cloud properties are constant along the line of sight.  The fraction of molecules in the lower state is given by
\be
\frac{N_r^l}{N_r}=\frac{1}{Z}e^{-E_l/kT_{\rm{ex}}}.
\ee
where $E_l$ is the energy of the ground state and $N_r$ is the total column density.

Now consider a population of CO molecules with isotope ratio $R\equiv n_{13}/n_{12}$, which in turn means that $N_{r13}=RN_{r12}$.  The ratio of the line intensities is then
\begin{multline}
\frac{I_{13}}{I_{12}}=\left(\frac{\nu_{13}}{\nu_{12}}\right)^2\left(\frac{A_{13}}{A_{12}}\right)\left(\frac{\sigma_{12}}{\sigma_{13}}\right)e^{-h(\nu_{13}-\nu_{12})/k_BT_{\rm{ex}}} \\ \times\left[\frac{1-\exp\left(-R\sigma_{13}N^l_{r12}\right)}{1-\exp\left(-\sigma_{12}N^l_{r12}\right)}\right],
\label{IR}
\end{multline}

The cross sections $\sigma_{12}$ and $\sigma_{13}$ corrected for stimulated emission are given by
\be
\sigma=\frac{3c^2A_{ul}}{8\pi \nu^2}\left(1-e^{-h\nu/k_BT_{\rm{ex}}}\right)\Delta\nu_{\rm{FWHM}},
\ee
where $\Delta\nu_{\rm{FWHM}}$ is the frequency full width at half maximum of the line \citep{Spitzer}.

The Einstein $A$ coefficients for the two transitions are $A_{1-0}^{^{12}\rm{CO}}=7.21\e{-8}$ s$^{-1}$ and $A_{1-0}^{^{13}\rm{CO}}=6.34\e{-8}$ s$^{-1}$ \citep{Muller2001}.  We assume that the molecular material has a constant excitation temperature $T_{\rm{ex}}=20$ K, and that the width of the line is dominated by a turbulent velocity distribution with a velocity FWHM of 10 km/s, similar to the velocities of the largest local GMCs \citep{1987ApJ...319..730S}, which translates to $\Delta\nu_{\rm{FWHM}}=3.83$ MHz for the $^{12}\rm{CO}$ line and 3.67 MHz for the $^{13}\rm{CO}$ line.  This yields cross sections $\sigma_{12}=3.67\e{-17}$ cm$^2$ and $\sigma_{13}=3.55\e{-17}$ cm$^2$, and
\be
\frac{I_{13}}{I_{12}}=0.84\frac{1-\exp\left(-0.34R\sigma_{13}N_{r12}\right)}{1-\exp\left(-0.36\sigma_{12}N_{r12}\right)},
\label{Iratio}
\ee
where the factors of 0.34 and 0.36 in the exponentials come from the conversion between $N_r^l$ and $N_r$.

\begin{figure}
\centering
\includegraphics[width=\columnwidth]{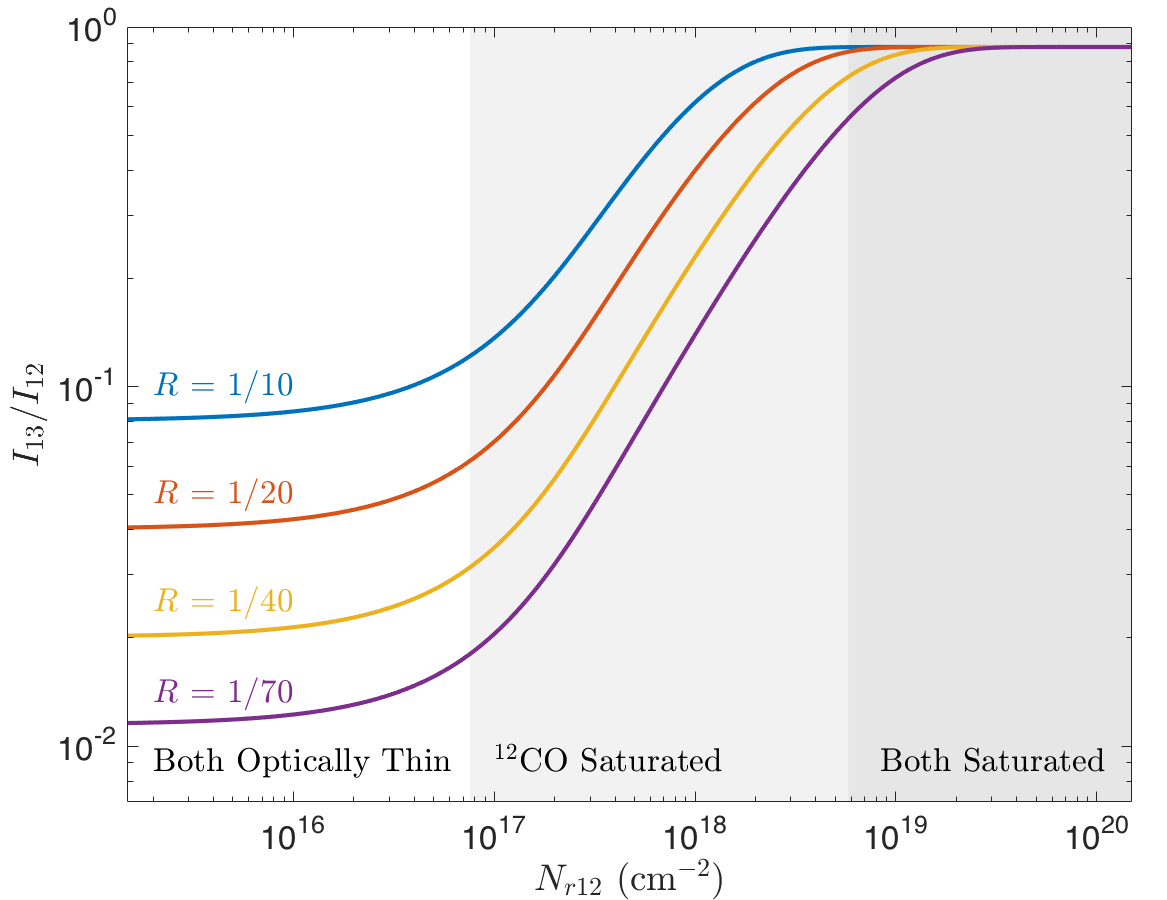}
\caption{Intensity ratios between the two different CO lines as a function of $^{12}$CO column density for different carbon isotope ratios.  The black dashed line shows the column density where the $^{12}$CO optical depth is unity, the colored dashed lines show where the $^{13}$CO optical depth is unity for the four shown values of $R$.}
\label{Ifig}
\end{figure}

If we expand this quantity in the limit of small optical depth, we find that $I_{13}/I_{12}$ goes to $0.8R$ at zero order.  In the opposite limit, where all of the emission is optically thick, we find that $I_{13}/I_{12}$ approaches 0.88.  Up to the differences in atomic-scale physics between the two CO species, this agrees with the prediction from Section 2.  Figure \ref{Ifig} shows the full behavior of this intensity ratio as a function of column density for different isotope ratios.  The two extremes can clearly be seen, along with a transition region where the $^{12}$CO line is optically thick but the $^{13}$CO line remains optically thin. Again, this result is independent on the detailed distribution of molecular material. We note, however, the precise values of the intensity ratio will be dependant on the excitation temperatures of the medium along the line of sight, as well as the details of selective chemistry and photodissociation. For the purpose of forecasting, however, we neglect these effects for simplicity. 

\subsection{Power spectra}
The primary observable for an intensity mapping survey is the power spectrum, which we compute assuming that the CO emission comes from a population of galaxies which are small compared to the survey resolution.  Each galaxy contributes a brightness temperature $T_b$ to the map, where the value of $T_b$ is drawn from a distribution $dn_{\rm{gal}}/dT_b$, which in turn depends on the galaxy luminosity function and the parameters of the intensity mapping instrument.  The total number density of emitting sources is then $n_{\rm{gal}}=\int dn_{\rm{gal}}/dT_b dT_b$.  

The three-dimensional auto-spectrum of an intensity map is typically written as
\be
P(k,z)=\left<T_b(z)\right>^2\overline{b}^2(z)P_m(k,z)+P_{\rm{shot}}(z),
\label{Aps}
\ee
where $P_m$ is the linear matter power spectrum, $\overline{b}$ is the luminosity-weighted galaxy bias, $\left<T(z)\right>$ is the sky-averaged intensity in the chosen line, here written in brightness temperature units, and $P_{\rm{shot}}$ is the scale-independent shot noise due to poisson fluctuations in the source positions (see, for example, \citet{Breysse2014}, \citet{Pullen2013}).  The mean intensity and shot noise can both be written in terms of the distribution of source intensities as
\be
\left<T_b\right>=\int_0^\infty T_b\frac{dn_{\rm{gal}}}{dT_b}dT_b,
\ee
\be
P_{\rm{shot}}=\int_0^\infty T_b^2\frac{dn_{\rm{gal}}}{dT_b}dT_b.
\ee

In addition to this auto-spectrum, we can compute a cross power spectrum for pairs of frequency bands separated by $\Delta\nu_{12/13}$, the difference in observed frequencies of the $^{12}$CO and $^{13}$CO lines at a desired redshift. The first term in Equation \ref{Aps} comes from correlations between pairs of sources, each weighted by the amount of intensity they produce.  The shot noise comes from the zero-lag correlation function, and thus essentially gives the correlation of each source with itself.  For the case of $^{12}$CO crossed with $^{13}$CO, the two lines are each tracing the same population of sources.  To compute the large-scale clustering term of a cross spectrum, we then simply need to weight one of each pair of sources by $^{12}$CO intensity and the other by $^{13}$CO intensity.  The cross spectrum is then
\begin{multline}
P_{12\times13}(k,z)=\left<T_{b,12}(z)\right>\left<T_{b,13}(z)\right>\overline{b_{12}}(z)\overline{b_{13}}(z)P_m(k,z)\\+P_{\rm{shot}}^{12\times13}.
\label{PkCross}
\end{multline}
If we assume the $^{13}$CO intensity is a function $T_{b,13}(T_{b,12})$ of the $^{12}$CO intensity, as implied by Figure 2, the new shot-noise term is given by
\be
P_{\rm{shot}}^{12\times13}=\int_0^\infty T_{b,12}T_{b,13}(T_{b,12})\frac{dn_{\rm{gal}}}{dT_{b,12}}dT_{b,12}.
\ee

Since we are considering cross-correlations between different frequency bands of an intensity map, it may be more intuitive to project down to two-dimensional power spectra, either within an individual band in the case of an auto-spectrum or between two bands in the case of a cross spectrum.  The angular power spectrum $C_\ell$ at a given multipole $\ell$ is given by
\begin{multline}
C_\ell = \frac{2}{\pi}\int k^2P(k)\left[\int f_1(r_1)j_\ell(kr_1)dr_1\right] \\ \times\left[\int f_2(r_2)j_\ell(kr_2)dr_2\right]dk,
\label{Cl}
\end{multline}
where $j_\ell$ is the spherical Bessel function of the first kind.  The selection functions $f_1$ and $f_2$ are set by the shape of the two frequency channels being correlated.  Equation (\ref{Cl}) is somewhat troublesome to evaluate exactly, so we will follow the approximations in Equations (2.21-22) of \citet{Breysse2014} to simplify our computations.

For the purposes of this work, we will assume that the selection functions $f_1$ and $f_2$ take the form of top-hat functions with widths set by the instrumental frequency bandwidths.  For a cross-correlation, these selection functions should be chosen so that $^{12}$CO emitters at redshift $z$ are included in $f_1$ and $^{13}$CO emitters at redshift $z$ are included in $f_2$, i.e. that $f_1$ and $f_2$ correspond to frequency bands separated by $\Delta\nu_{12/13}$.  From here on we will assume for simplicity that the redshift ranges of $f_1$ and $f_2$ overlap exactly.  This may not be the case for a real experiment, but this will not alter our final results.  Note also that $f_1$ will contain $^{13}$CO emitters and $f_2$ will contain $^{12}$CO emitters from outside the target redshift range.  This should not be an issue, as the spatial separation between the $^{12}$CO emitters in the two bands is large enough that they should be uncorrelated on the scales we care about.  See Appendix A for a quantitative derivation of this result.

\section{Forecast}
In the previous section we outlined the basic physical process which controls the emission of these two lines within a cloud and the final observables which will be computed from an intensity map.  What we need now to forecast constraints on these observables is a model connecting the cloud-scale physics to the cosmological-scale power spectra.  Given the large astrophysical uncertainties at these high redshifts, we will model this connection in a fairly simple manner.  However, the advantage of the formalism presented above is that it holds regardless of exactly how the molecular gas is modeled.  One could easily replace the calculations presented below with a more sophisticated model if one were made available.

At the redshifts we are considering, virtually all CO emission comes from within galaxies.  Since our beam is large compared to an individual galaxy, we can model each galaxy as a point source with a given $L_{12}$ and $L_{13}$.  We can predict the ratio of these luminosities by slightly modifying Equation (\ref{Iratio}):
\be
\frac{L_{13}}{L_{12}}=0.88\frac{1-\exp\left(-0.34R\sigma_{13}\overline{N}_{r12}\right)}{1-\exp\left(-0.36\sigma_{12}\overline{N}_{r12}\right)},
\label{Lratio}
\ee
where $\overline{N}_{r12}$ is now the $^{12}$CO column density averaged over all lines of sight through the galaxy.  These luminosities can then be converted to brightness temperature by
\be
T_b = \frac{c^3}{8\pi k_B \nu_{\rm{obs}} H(z)}\frac{L}{V_{\rm{vox}}(1+z)},
\label{LtoT}
\ee
where $k_B$ is Boltzmann's constant, $H(z)$ is the Hubble parameter, and $V_{\rm{vox}}$ is the comoving volume of an individual map voxel.

Equation (\ref{Lratio}) relates the luminosity of a galaxy in the two CO lines to the properties of its molecular clouds.  In order to compute power spectra from a population of such galaxies, we will use empirical results to compute a distribution of $^{12}$CO luminosities, as well as a relation between $^{12}$CO luminosity and molecular cloud density.  With these tools in hand, we can predict a distribution of $^{13}$CO luminosities and forecast a power spectrum for an intensity mapping experiment.

\subsection{$^{12}$CO Luminosity Function}
The literature contains a wide variety of methods for modeling CO luminosity functions at high redshift, ranging from simple scaling arguments \citep{Visbal2010,Pullen2013} to sophisticated semianalytic calculations \citep{Lagos2012,Popping2016}.  We are more interested here in the relation between $^{12}$CO and $^{13}$CO rather than the exact behavior of the $^{12}$CO line, so we will err on the side of simplicity.  We assume that the $^{12}$CO luminosity function takes the form 
\be
\frac{dn}{dL_{12}}=\phi_*\left(\frac{L_{12}}{L_{*}}\right)^\alpha \exp\left(-\frac{L_{12}}{L_*}-\frac{L_{\rm{min}}}{L_{12}}\right),
\label{LF}
\ee
which is simply a Schechter function with an exponential cutoff added to the low-luminosity end.  This choice has the advantage that a wide variety of CO emission models can be expected to produce luminosity functions with similar shapes.

We choose values for the free parameters $(\phi_*,\alpha,L_*,L_{\rm{min}})$ by comparing to the results of \citet{Li2016}.  This model uses a simulated relationship between halo mass and SFR from \citet{Behroozi2013}, then uses scaling relationships between SFR and far infrared luminosity and between FIR luminosity and CO luminosity to get $L_{12}$ as a function of halo mass.  Finally, they add lognormal scatter around this relation to obtain a luminosity function.  If we fit our Schechter function to these results, we get best fit parameters $\phi_*=2.8\times10^{-10}\ \left(\rm{Mpc}/h\right)^{-3}\ L_{\sun}^{-1}$, $\alpha=-1.87$, and $L_*=2.1\times10^6\ L_{\sun}$.  \citet{Li2016} assume a hard cutoff in $^{12}$CO luminosity for halos smaller than $10^{10}\ M_{\sun}$.  We set the location of our low-luminosity cutoff at $L_{\rm{min}}=500\ L_{\sun}$, which is the luminosity in their model which corresponds to $10^{10}\ M_{\sun}$.

In order to compute power spectra from Equations (\ref{Aps}) and (\ref{PkCross}), we also need a prediction for the luminosity-weighted bias $\overline{b}$.  In principle, the luminosity weighting means that the bias should take into account the full luminosity function model.  However, in the interest of keeping things simple, we will follow \citet{Gong2014} in assuming for the purpose of the bias calculation that luminosity is proportional to halo mass.  The bias is then given by
\be
\overline{b}=\frac{\int b(M) M dn/dMdM}{\int M dn/dMdM}.
\label{bias}
\ee
We use the \citet{Tinker2008} mass function $dn/dM$ and the corresponding mass-dependent bias $b(M)$ from \citet{Tinker2010}.  When calculating the cross spectrum between $^{12}$CO and $^{13}$CO we will use the same bias factor for both lines.  One could just as easily assume a more sophisticated weighting in Equation (\ref{bias}) and adjust it slightly to take into account a relation between $L_{12}$ and $L_{13}$.  Since the two lines trace the same population of galaxies, this would cause at most an order unity change in the final amplitudes of our power spectra, and likely much smaller.  Using the \citet{Li2016} $L(M)$ relation to weight the bias, for example, only increases $\overline{b_{12}}$ by $\sim5\%$.  This is a very small effect, especially since the astrophysical uncertainties in the rest of the model are so very large.

\subsection{Relating Luminosity to Column Density}
In order to determine the distribution of $^{13}$CO intensities, we would like to rewrite the ratio $L_{13}/L_{12}$ from Equation (\ref{Lratio}) in terms of the $^{12}$CO luminosity $L_{12}$.  Doing so requires a relationship between average column density $\overline{N}_{r12}$ and $L_{12}$, which we will derive here using a similar set of scaling relations to those used to get the \citet{Li2016} model described above.  

We start with the assumption that the average ratio $Z_{\rm{CO}}$ of $^{12}$CO and H$_2$ column densities is roughly $10^{-4}$, following \citet{Bolatto2013} based on the observations of \citet{Sofia2004}.  We can then convert to a mean surface star formation rate density $\Sigma_{SFR}$ using the Schmidt-Kennicutt law
\begin{multline}
\frac{\Sigma_{SFR}}{\SigSFRunits}=2.5\e{-4}\left(\frac{\Sigma_{\rm{H}_2}}{M_{\sun}\ \rm{pc}^2}\right)^{1.4}\\=4.8\e{-32}\left(\frac{N_{\rm{H_2}}}{\rm{cm}^{-2}}\right)^{1.4},
\end{multline}
where we have assumed that the mass of the gas is dominated by H$_2$.  The star formation rate is then simply
\be
SFR=\pi r_{\rm{gal}}^2\Sigma_{SFR},
\ee
where we assume a representative galaxy radius $r_{\rm{gal}}=30$ kpc for all galaxies.  We relate star formation rate to $L_{12}$ using Model A of \citet{Pullen2013}, which uses the relation between SFR and far infrared luminosity from \citet{Kennicutt1998} and the relation between FIR luminosity and CO luminosity from \citet{Wang2010} to obtain
\be
\frac{L_{12}}{L_{\sun}}=3.2\e4\left(\frac{SFR}{M_{\sun}\ \rm{yr}^{-1}}\right)^{3/5}.
\ee
Combining all of this together yields
\be
\frac{\overline{N}_{12}}{\rm{cm}^{-2}}=X_L\left(\frac{L_{12}}{L_{\sun}}\right)^{1.2},
\label{LtoN}
\ee
where $X_L\equiv3.5\e{10}$.  The luminosity ratio of our two CO lines is then
\be
\frac{L_{13}}{L_{12}}=0.84\frac{1-\exp\left[-0.34R\sigma_{13}X_L\left(L_{12}/L_{\sun}\right)^{1.2}\right]}{1-\exp\left[-0.36\sigma_{12}X_L\left(L_{12}/L_{\sun}\right)^{1.2}\right]},
\label{Lratio2}
\ee
These luminosities then can be converted to intensities using Equation (\ref{LtoT}).

\subsection{Experimental Parameters}
When forecasting the constraining power of the measurements discussed here, we will consider three different surveys: the COMAP ``Pathfinder" and ``Full" surveys described in \citet{Li2016} and a hypothetical ``Future" survey which has improved sensitivity, resolution, and additional observing time.  The parameters we use for these three instruments are given in Table \ref{ParamTable}.  Values for the ``Pathfinder" and ``Full" experiments are obtained from Table 2 of \citet{Li2016}.  Construction on the ``Pathfinder" survey is currently underway, the ``Full" experiment is planned as a next step.  The ``Future" experiment assumes a modest increase in sensitivity and angular resolution, and either additional observing time or an array of several dishes observing simultaneously.

\begin{table}
\centering
\caption{Survey parameters used for Fisher analysis.  Values for the ``Pathfinder" and ``Full" experiments are obtained from Table 2 of \citet{Li2016}}
\begin{tabular}{cccc}
\hline
Parameter & Pathfinder & Full & Future \\
\hline
Frequency range (GHz) & 30-34 & 30-34 & 30-34 \\
Patch Area $\Omega_s$ (deg$^2$) & 2.5 & 6.25 & 30 \\
Beam Size $\theta_{\rm{FWHM}}$ (arcmin) & 6 & 3 & 2 \\
Observing time/Patch $t_{\rm{obs}}$ (hr) & 1500 & 2250 & $2250\times5$\\
Number of patches $N_{\rm{patch}}$ & 4 & 4 & 4 \\
Sensitivity $s$ ($\mu$K s$^{1/2}$) & 1026 & 783 & 585 \\
Channel width $\Delta\nu$ (MHz) & 40 & 10 & 10 \\
Number of channels $N_{\rm{ch}}$ & 100 & 400 & 400 \\
\hline
\end{tabular}

\label{ParamTable}
\end{table}

\subsection{Fisher analysis}
\begin{figure}
\includegraphics[width=\columnwidth]{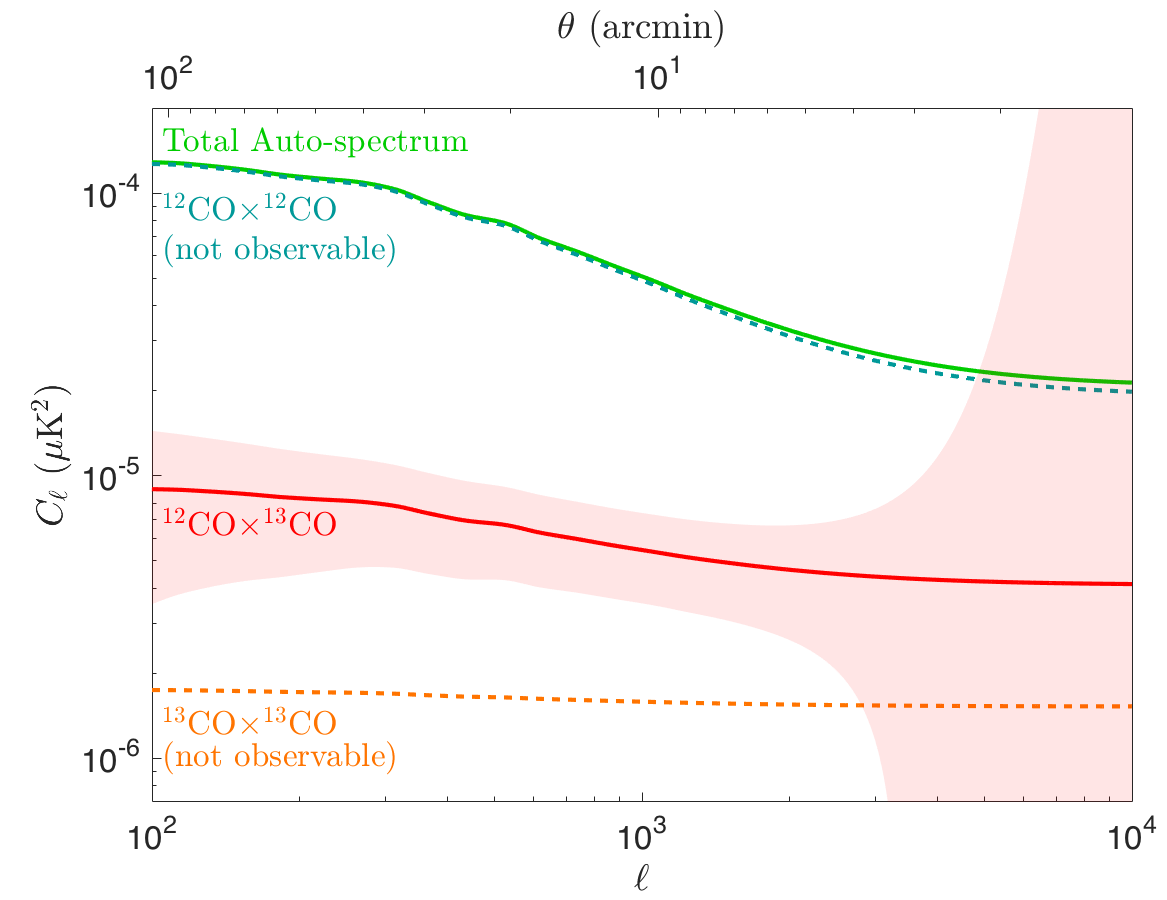}
\caption{Angular power spectra for different combinations of the two CO lines.  Dashed lines show auto-spectra of $^{12}$CO (blue) and $^{13}$CO (orange), which are not independently observable.  Solid lines show the two observable spectra, the auto-spectrum of a single frequency band (green) and the cross-spectrum of two bands separated by $\Delta\nu_{12/13}$ (red).  The red shaded region shows the instrumental error on the cross-spectrum assuming the parameters of the ``Full" experiment and multipoles binned in sets of 10.}
\label{Cls}
\end{figure}

Combining the results from previous sections assuming the frequency bandwidth of the ``Full" or ``Future" experiments yields the four power spectra shown in Figure \ref{Cls}.  We assume a fiducial CO isotopologue ratio $R=1/70$, which is representative of the many local and extragalactic measurements (for example, \citealt{Wilson1999}, \citealt{Davis2014}, \citealt{Smith2015})

  The blue and orange dashed curves show the auto-spectra for the $^{12}$CO and $^{13}$CO lines respectively.  Neither of these are observable on their own, as correlating any given band with itself will simply give the sum of the two, which is shown in green.  This total auto-spectrum is essentially just the $^{12}$CO spectrum, with a correction of order a few percent from the $^{13}$CO emitters.  The red curve shows the cross spectrum between two bands separated by $\Delta\nu_{12/13}$.  Note that the quantity plotted on the $y$-axis of Figure \ref{Cls} is $C_\ell$ rather than the commonly seen $\ell(\ell+1)C_\ell/(2\pi)$.

The instrumental noise from a given instrument can be modeled as an additional Gaussian random field on the sky with power spectrum
\be
C_\ell^N=\frac{\Omega_s s^2}{t_{\rm{obs}}}\exp\left[\frac{\theta_{\rm{FWHM}}^2\ell(\ell+1)}{16\ln2}\right],
\ee
where $\Omega_s$ is the solid angle of a given patch, $s$ is the sensitivity in $\rm{\mu}$K s$^{1/2}$, $t_{\rm{obs}}$ is the amount of time spent observing a given patch, and $\theta_{\rm{FWHM}}$ is the beam full width at half maximum \citep{Tegmark1997}.  The uncertainty of a given $C_\ell$ is then
\be
\sigma_\ell=\sqrt{\frac{8\pi}{N_{\rm{ch}}N_{\rm{patch}}\Omega_s(2\ell+1)}}\left[C_\ell^N+C_\ell\right],
\ee
where $N_{\rm{ch}}$ is the number of frequency channels stacked in $N_{\rm{patch}}$ sky patches \citep{Jaffe2000}.  The second term in parenthesis gives the sample variance error from patch to patch.  The shaded region in Figure \ref{Cls} shows the error on the cross-spectrum assuming the parameters of the ``Full" experiment with multipoles binned in sets of 10.

Note that for these calculations we are treating an intensity map in a given patch as a stack of independent two-dimensional maps from each frequency bin.  This approach neglects line-of-sight Fourier modes.  As stated in \citet{Pullen2013}, this means that our simplified approach likely underestimates the constraining power of a given instrument somewhat.  In addition, we have assumed here that the properties of the two observed lines do not vary significantly across the observed redshift range.  We expect errors introduced due to these approximations to be small compared to the large uncertainties present in the modeling of these spectra.

We stated previously that the cross-spectrum of these two CO lines is interesting because it depends on both the relative abundances of the two carbon species and the distribution of molecular gas densities at a given redshift.  The abundance information can be trivially parameterized using the isotopologue ratio $R$, which depends on the nucleosynthesis history.  We parameterize the gas density distribution by defining the quantity
\be
f_s\equiv\frac{\int_{T_\tau}^\infty T_{b,12} dn_{\rm{gal}}/dT_{b,12}dT_{b,12}}{\int_{0}^\infty T_{b,12} dn_{\rm{gal}}/dT_{b,12}dT_{b,12}},
\label{fs}
\ee
which is the fraction of the measured $^{12}$CO emission coming from optically thick lines of sight.    In order to constrain $f_s$, we treat the quantity $X_L$ in Equation (\ref{LtoN}) as a free parameter.  The $^{12}$CO line becomes optically thick when $\overline{N}_{12}\sigma_{12}=1$, or when $\overline{N}_{12}=6\times10^{16}$ cm$^{-2}$.  From a given value of $X_L$ we can compute an intensity $T_\tau$ from Equations (\ref{LtoN}) and (\ref{LtoT}) which corresponds to this column density, then use Equation (\ref{fs}) to compute a value of $f_s$.  Our fiducial value for $X_L$ corresponds to a saturated fraction $f_s=0.36$.

No relevant constraints on these quantities currently exist at these redshifts.  The parameters $R$ and $f_s$ as we have defined them here are values averaged over all of the molecular gas in a given redshift slice.  Since only an intensity mapping survey can access the vast majority of the galaxy population, these quantities cannot be realistically constrained by existing data.  This also means that even weak constraints from an intensity map hold great scientific value.

\begin{figure}
\centering
\includegraphics[width=\columnwidth]{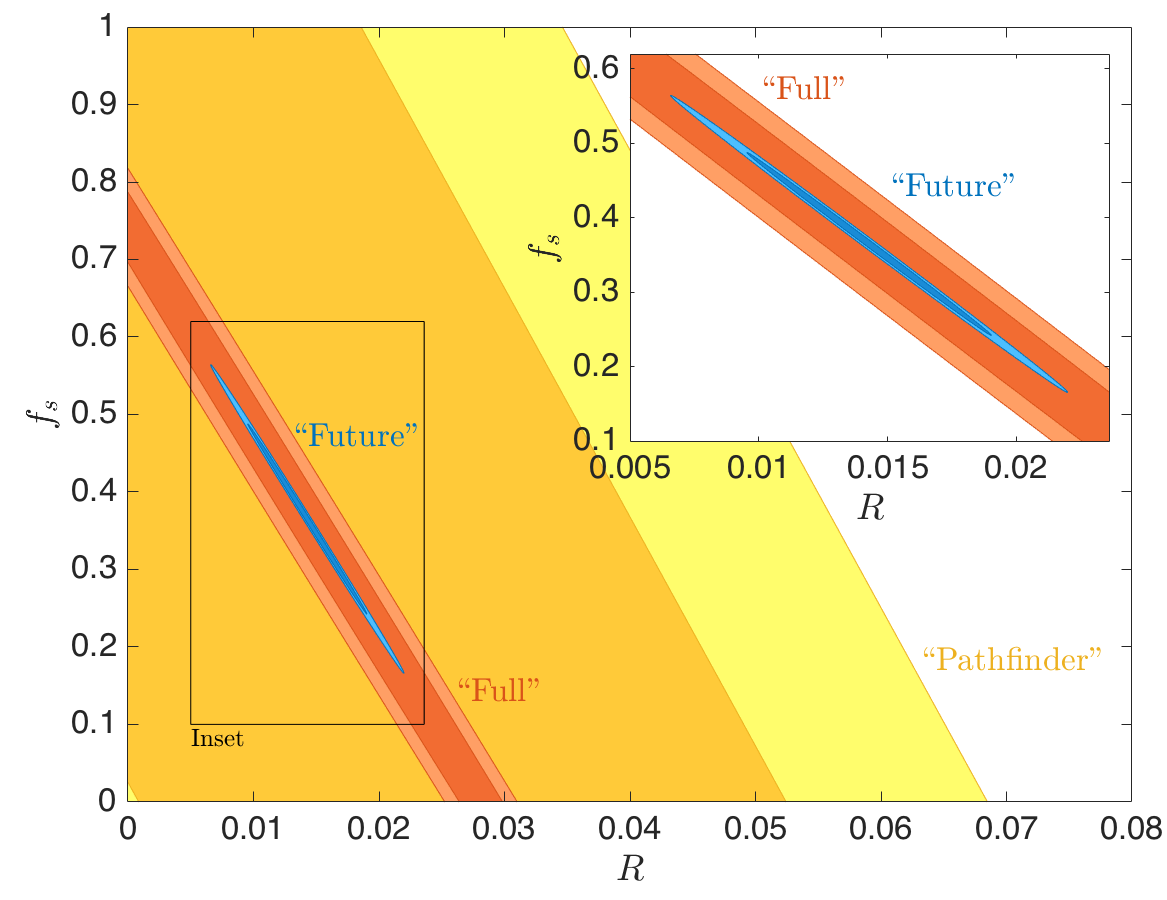}
\caption{Forecasted 1- and 2-sigma constraints on isotopologue ratio $R$ and saturated fraction $f_s$ for the ``Pathfinder" (yellow), ``Full" (red), and ``Future" (blue) experiments.}
\label{Fisher}
\end{figure}

Figure \ref{Fisher} shows the results of our Fisher analysis for our three experiments.  Dark colors show 1-$\sigma$ constraints, light colors show 2-$\sigma$ constraints.  The ``Pathfinder" survey only yields a signal-to-noise ratio of $\sim1$ for the cross-spectrum, so it effectively provides an upper limit on the cross-correlation signal.  The ``Full" survey does better, however the constraints on the two parameters are quite degenerate, leading to order unity 1-$\sigma$ fractional errors on both parameters.  Despite this degeneracy, the volume of parameter space is still dramatically reduced compared to the complete lack of constraints currently available.  Fractional errors on our two parameters fall to $\sim30\%$ for the ``Future" experiment.

For these constraints, we have assumed that the $^{12}$CO luminosity function is known exactly.  This may seem like a poor assumption, since the cross-spectrum depends equally on the properties of both lines.  However, the auto-spectrum, which is essentially the $^{12}$CO spectrum, will be measured as well.  The auto-spectrum has an amplitude roughly 10 times greater than the cross-spectrum, so the remaining uncertainty in the cross-spectrum from $^{12}$CO emission should be very small compared to that from $^{13}$CO emission, especially if things like the one-point PDF of the map \citep{Breysse2016} are used to further constrain $dn/dL_{12}$.  For more detail, see Appendix \ref{12marg}.

\section{Discussion}
Using intensity mapping, we have demonstrated that it is possible to constrain both carbon isotopologue ratios and molecular gas density distributions at cosmological distances.  These measurements cannot be made using any other method, as even the deepest targeted survey would only be able to study a handful of the brightest CO emitters, as illustrated in Figure 8 of \citet{Li2016}.  Intensity mapping surveys will allow measurement of these quantities from the entire galaxy population.  Using the cross-correlation techniques discussed here also allows significant improvement beyond what is possible using intensity mapping of only a single spectral line, allowing for a much more in-depth probe of the molecular gas which feeds star formation.  

As seen in Figure 4, constraints on the parameters $R$ and $f_s$ are quite degenerate for all three of our surveys.  This degeneracy arises because, as shown in Figure \ref{Cls}, most of the signal-to-noise comes from the clustering component of the signal, which depends only on the mean intensities of the two lines.  Using the mean intensity alone cannot distinguish between a higher $^{13}$CO abundance or more saturated $^{12}$CO emission.  This degeneracy can be broken somewhat by a measurement of the shot noise.  Indeed, a large part of why the ``Future" survey results in a weaker degeneracy is because it has much more constraining power in the shot-noise component of the spectrum.  It may thus be possible to reproduce some of the extra constraining power of the ``Future" experiment by combining the ``Full" experiment with an interferometric intensity mapping experiment.  Interferometers such as the one used in the COPSS \citep{Keating2016} survey are sensitive to considerably smaller scales than single-dish COMAP-like experiments.  This makes them well suited to studying the shot-noise dominated regime.

One could also potentially further improve these constraints by combining an intensity map with traditional galaxy survey data.  Even if only the brightest end of the luminosity function were constrained through these surveys, that could still help to break this degeneracy.  An in-depth study of quasar lines could accomplish the same purpose.  Neither of these measurements can hope to do as well as our intensity mapping forecasts on its own, but they could provide valuable complements.  If the degeneracy between $R$ and $f_s$ were completely broken, the fractional uncertainties on these quantities in our forecast for the ``Full" survey would drop from order unity to a few percent.  It is probably overly optimistic to assume perfect degeneracy breaking, but the potential for dramatic improvement is clear.

For the above forecasts, we constrained the properties of the CO emission averaged over the entire survey volume.  However, for a sufficiently sensitive survey, one could break the volume down into multiple redshift bins and constrain the redshift variation of these quantities.  One could then compare to other redshift-dependent measurements, such as the star-formation history given in \citet{Madau2014}.  We used a ``Future" survey targeting the same redshift range as the planned COMAP ``Pathfinder", but there is no reason why such a futuristic experiment could not be extended to a broader redshift range.  In addition, some have proposed conducting CO intensity mapping surveys at $z\sim6-10$ as a probe of reionization physics \citep{Lidz2011}.  At these redshifts, we begin to run into the limits of what is known about molecular gas, the formation of carbon and consequently CO, and the formation of the earliest generations of stars where metallicity is limited.  With sufficient sensitivity, one could apply the cross-correlation methods applied here to understanding this mysterious era in cosmic history.

It is important to note here that the forecasts given in this work are based on deliberately simplified modeling.  We have neglected a number of effects here which deserve more consideration in future work.  For example, our model assigns a $^{13}$CO luminosity to a given galaxy based entirely on its average $^{12}$CO column density, which in turn is determined solely by its $^{12}$CO luminosity.  In reality, a galaxy will contain a large number of heterogeneous molecular clouds with different densities, leading to a more complicated relation between the two CO lines.  In the short term, this could perhaps be taken into account by simply assuming a scatter in the luminosity relations, similar to what was done in \citet{Li2016}.  The basic framework we have outlined here can be straightforwardly generalized to more sophisticated models as we gain a better understanding of the environments in these distant galaxies.

The interpretation of this type of measurement may also be complicated by the fact that the metallicity of high-redshift systems could deviate significantly from solar.  It is well-established that the molecular gas to CO ratio increases in lower-metallicity systems \citep{Bolatto2013}.  This fact must be taken into account when attempting to constrain global molecular gas properties from these observations, and will add additional uncertainty.  In addition, low metallicity may alter the correlation between molecular gas and star formation.  In such systems, star formation can proceed in regions which remain dominated by atomic gas.  These regions will produce much less CO emission than would be expected from star formation at solar metallicity \citep{Krumholz2012,Glover2012}.

We have also neglected the effects of various foregrounds in this analysis.  However, we do not expect this to have a substantial impact on the results in this case.  There are two types of foregrounds which affect intensity mapping surveys.  The first are those with continuum spectra, such as Milky Way dust and synchrotron emission.  These foregrounds have been extensively studied for the case of 21 cm reionization experiments (\citet{Morales2010} and references therein), which have a considerably higher foreground-to-signal ratio than CO experiments.  Since these foregrounds have little spectral structure, they only contribute substantially to Fourier modes which fall close to the line-of-sight, and can thus be cleaned simply by subtracting these modes from an analysis.  The second type of foreground are spectral lines from other redshifts which fall into our target frequency bands.  As mentioned previously, $^{13}$CO can be thought of as a line foreground to $^{12}$CO, though not an important one.  A sufficiently bright line can in principle contaminate the CO auto-spectrum \citep{Breysse2015}, but unless there exists a pair of foreground lines separated by exactly $\Delta\nu_{12/13}$ there should be no contamination to the cross-spectrum we looked at here.

As mentioned previously, intensity mapping surveys are planned in many other lines besides CO, such as [CII], Ly$\alpha$, and the 21 cm HI line.  The broad strokes of the work we have presented here could easily be applied to cross-correlation between CO and these other lines.  Such cross-correlations would require additional planning compared to our CO isotope correlation, since both lines would not appear in the same survey.  However, if experiments were planned well to target the same volumes, one could potentially learn about high-redshift galaxies in even greater detail, allowing us to study the complex ecology of gas dynamics and star formation across the entire history of the universe.

\section{Conclusion}
We have demonstrated here a method whereby we can dramatically improve our understanding of molecular gas at high redshifts by combining information from $^{13}$CO in intensity maps with the usual $^{12}$CO.  This is a direct extension of similar techniques used when studying local molecular clouds.  By cross-correlating properly chosen slices of a CO intensity map it is possible to determine the total amount of $^{13}$CO emission and how it varies with $^{12}$CO emission.  We showed how the cross-spectrum of these two lines can be used to constrain not only the abundance ratio of these two species but also the density distribution of molecular gas in the mapped galaxy population, quantities which are extremely difficult if not impossible to measure with any other method.  This technique will allow us to gain deep insights into the processes that feed star formation throughout cosmic history.  By branching out to more detailed models, additional spectral lines, and broader redshift ranges we can study the complex ecology of star formation and galaxy evolution at a level of detail unimaginable with traditional methods.  We have likely only scratched the surface of what intensity mapping can teach us about the distant universe.

The authors would like to thank Ely Kovetz, Marc Kamionkowski, Tony Li, Christopher Matzner, Julia Roman-Duval for useful discussion.  The authors would also like to thank the referee for helpful comments which improved the paper.  PB was supported by the Simons Foundation.

\appendix
\section{Spurious $^{12}$CO Correlation}\label{12corr}
When we take the cross-spectrum between two chosen bands, our goal is to get at the correlation between $^{12}$CO and $^{13}$CO emission from galaxies at a single redshift.  However, the cross-spectrum also contains power from pairs of $^{12}$CO emitters and pairs of $^{13}$CO emitters at different redshifts.  For example, cross-correlating bands centered at 32 and 30.6 GHz would correlate emission from $^{12}$CO and $^{13}$CO at $z=2.6$.  It will also correlate $^{12}$CO emission from $z=2.6$ and 2.8 and $^{13}$CO emission from $z=2.4$ and 2.6.  If the spatial separation between these pairs of redshifts is large compared to the scale set by a given multipole, then this spurious correlation should be small.  Here we demonstrate this fact quantitatively.

The 3D auto spectrum $P_{\rm{CO}}(k,z)$ of a CO line is given by Equation (\ref{Aps}).  If we wish to compute the angular cross-spectrum between two different redshift bands we simply apply Equation (\ref{Cl}) with the $^{12}$CO 3D spectrum and selection functions $f_1$ and $f_2$ centered on our two chosen bins.  As stated in \citet{Breysse2014}, we can evaluate this expression in two limits.  If our bands have a comoving spatial width $\delta r$ which satisfies $\ell \delta r/r\gg 1$ (i.e. if we consider fluctuations on length scales small compared to the width of the redshift slice), we can use the well-known Limber approximation
\be
C^{12\times12,s}_\ell\approx\int\frac{H(z)}{c}\frac{f_1(z)f_2(z)}{r^2(z)}P_{\rm{CO}}\left[k=\ell/r(z),z\right]dz,
\ee
\citep{Limber1953,Rubin1954}.  In this limit, it is clear that if $f_1$ and $f_2$ do not overlap for any value of $z$, this integral vanishes and we will see no spurious correlation in our cross-spectrum.  This agrees with our previous intuition, as the Limber approximation is valid for large $\ell$'s where we expected our signal to be small.  

In the opposite limit, where $\ell\delta r/r\ll1$, we can replace both selection functions with Dirac delta functions centered at the two redshifts $z_1$ and $z_2$.  Evaluating Equation (\ref{Cl}) in this limit gives
\be
C^{12\times12,s}_\ell\approx\frac{2}{\pi}\int k^2P_{\rm{CO}}(k)j_\ell\left[kr(z_1)\right]j_\ell\left[kr(z_2)\right]dk.
\ee
Since the extra $^{12}$CO correlation vanishes in the Limber approximation, we can take this narrow-band approximation as an upper limit on the amount of spurious power.  If we evaluate this integral numerically at $\ell=100$, we find a value for $C^{12\times12,s}_\ell$ which is approximately 1\% of the cross-spectrum $C^{12\times13}$.  The ratio falls to $\sim$ a few parts in $10^{5}$ at $\ell=500$.  The largest scales accessible to the instruments we consider here are around $\ell\sim100$, so we can safely ignore this extra power in our analysis.  Since the $^{13}$CO line is so much fainter than the $^{12}$CO line, spurious $^{13}$CO correlation will be even less significant.

It should be noted that this argument would not hold if we were to consider an experiment with a substantially larger survey area.   At $\ell=10$, contamination from this extra $^{12}$CO in the narrow-band approximation rises to $\sim60\%$ of the cross-spectrum.  This is still an overestimate, but it would still likely need to be taken into account if performing this analysis on very large scales.  It may also needed to be taken into account for measurements at higher redshifts, such as those which would target the Epoch of Reionization \citep{Lidz2011}.  As the target redshift increases, the comoving separation of the two $^{12}$CO populations in the two bands decreases, leading to additional correlation on smaller scales.  This is particularly significant because these band pairs in a survey at $z=7$ will cover populations of $^{12}$CO emitters separated by nearly the baryon acoustic oscillation scale, which will significantly boost the correlation.

\section{Marginalizing over $^{12}$CO}\label{12marg}
\label{12marg}
For the molecular gas constraints shown in Figure \ref{Fisher}, we assumed that the $^{12}$CO luminosity function was known exactly.  Here we will relax that assumption and show that the effects on our final constraints is small.  Unfortunately, we cannot simply perform a full Fisher analysis combining all four $^{12}$CO parameters from Equation (\ref{LF}) with the molecular gas parameters $R$ and $f_s$.  Even if we use both the auto- and cross-spectra, the only degrees of freedom we can use are the amplitudes of the clustering and shot noise components of each spectrum.  This leaves us with four degrees of freedom and six parameters, ensuring that several parameter constraints will be perfectly degenerate.

This issue arises due to the fact that the power spectrum only contains all of the information about a map if that map is purely Gaussian \citep{Peebles1980}.  Since a galaxy's CO luminosity is determined by a variety of nonlinear astrophysical processes, the intensity distribution in an intensity map will be very non-Gaussian.  In order to fully constrain the luminosity functions which give rise to these non-Gaussian maps, we need some prior information from another source.  One promising source of extra information is the one-point pixel intensity distribution.  \citet{Breysse2016} showed that this statistic could be used with an experiment similar to the COMAP ``Full" survey to constrain the $^{12}$CO luminosity function to within $\sim10-20\%$.

Consider then a Fisher matrix computed from both power spectra over all six parameters.  We can add to this a prior matrix assuming uncorrelated 10\% errors on the four $^{12}$CO parameters.  The new Fisher matrix will then be invertible, allowing us to forecast parameter constraints.  If we perform this analysis, the fractional 1-$\sigma$ error on $R$ increases to $\sim$20\% from the $\sim$15\% value quoted above.  As stated above, this increase is small because the $^{12}$CO emission is substantially brighter that the $^{13}$CO emission.  Uncertainties on $R$ and $f_s$ will therefore always be dominated by uncertainty in the $^{13}$CO measurements.

\end{document}